\documentstyle[prb,%preprint,
aps,psfig]{revtex}
\begin{document}
\draft

%\onecolumn

\title{Density functional study of Ni bulk, surfaces and the adsorbate
systems Ni(111)
$(\protect\sqrt{3} \times \protect\sqrt{3})$R30$^\circ$-Cl, and
Ni(111)${(2 \times 2})$-K}
\author{K. Doll}
\address{Institut f\"ur Mathematische Physik, TU Braunschweig,
Mendelssohnstra{\ss}e 3, D-38106 Braunschweig}
\maketitle

\begin{abstract}
Nickel bulk, the low index surfaces and the adsorbate systems
Ni(111)
$(\protect\sqrt{3} \times \protect\sqrt{3})$R30$^\circ$-Cl, and
Ni(111)${(2 \times 2)}$-K are studied with gradient corrected
density functional calculations. It is demonstrated that an approach based
on Gaussian type orbitals is capable of describing these systems.
The preferred adsorption sites and  geometries are in good agreement
with the experiments. Compared to non-magnetic substrates, there
does not appear to be a huge difference concerning the structural
data and charge distribution. The magnetic moment of the nickel
atoms closest to the adsorbate is reduced, and oscillations of the 
magnetic moments
within the first few layers are observed in the case
of chlorine as an adsorbate. The trends observed for the
Mulliken populations of the adsorbates are consistent 
with changes in the core levels.
\end{abstract}

\pacs{ }

\narrowtext
\section{Introduction}

Adsorption on surfaces is one of the main focuses in surface 
science due to the enormous importance of topics 
such as corrosion or catalysis. A huge number of experiments has
been performed to understand these processes.
In addition, theoretical studies have become capable of 
modelling the adsorption process. 

The adsorption 
of alkali metals or haloges on metallic surfaces may be considered
as two prototypes of adsorption processes. They are
fairly simple and experimentally well studied. However, there
have been quite a few surprises. For example, the occupancy
of top sites by alkali metals as adsorbates was unexpected (for reviews, 
see references \onlinecite{DiehlMcGrath,Over,StampflScheffler} and more 
general, reference \onlinecite{Zangwill}). 
Halogens are another type of adsorbate which is experimentally
well studied. On close-packed
(111) surfaces, they usually occupy the face centered cubic (fcc)
hollow site, and in a few systems a partial occupancy (around
15-25\%) of
hcp (hexagonal close-packed) sites has also been observed
\cite{Takata1992,Kadodwala}. In addition, there
is now a growing set of simulations for halogens on
metallic surfaces (e.g. chlorine 
\cite{DollHarrison2000,DollHarrison2001,Jia2003},
bromine \cite{Wang2002,Blum2002} or iodine \cite{YunWang2001}).
Both for alkali metals and for halogens as adsorbates, there are
general questions such as the geometry, binding energies, diffusion
barriers, charge transfer and the binding mechanism.

Magnetic surfaces pose additional difficulties and questions such
as how the magnetic moment changes after adsorption. It is
thus an interesting question which insights  simulation can 
give for these systems. Nickel surfaces and adsorption thereon
have been well studied by experimental groups and thus are a
prototype system for a computer simulation. For example,
hydrogen on Ni(111) \cite{Kresse2000}, benzene on Ni(111)
 \cite{Mittendorfer2001},
CH$_3$ dehydrogenation on Ni(111) \cite{Michaelides2000}
or CO on Ni(110)\cite{Geetal2000} adsorption
have been simulated recently. The present article will deal with the
simulation of alkali and halogen adsorption, with Cl and K as
examples.

The intention of this article is to give a broad overview over nickel
and the adsorbate systems. The computational technique 
is based on a local basis
set of Gaussian type functions. As nowadays most calculations,
especially for metals, rely on plane-waves, the following section 
\ref{Nibulksection} was devoted
to the fairly simple systems of Ni bulk and the low index surfaces.
This way, it will be demonstrated that results from a code based
on Gaussian type orbitals reproduce
the ones from plane-wave codes. In sections \ref{chlorineadsorptionsection}
and \ref{potassiumadsorptionsection},
the results for chlorine and potassium adsorption are presented, 
respectively. Finally, the results are summarized and discussed.
Additional details on the calculations are given in the appendix.

\section{Ni bulk and low index surfaces}
\label{Nibulksection}

\subsection{Computational Parameters}

A local basis set formalism was employed where the basis functions are
Gaussian type orbitals centered at the atoms as implemented in the code
 CRYSTAL \cite{Manual}. The calculations
were done with spin-polarization, using the gradient corrected functional of
Perdew, Burke and Ernzerhof (PBE) \cite{PBE}, and in some additional
calculations the gradient corrected functional by Perdew and Wang (PW)
\cite{Perdewetal}. These functionals perform actually nearly identical
(see, e.g. references 
\onlinecite{PBE,Hammeretal}, or with the CRYSTAL code for Ag
\cite{DollHarrison2001}), and also in the present article the results
with both functionals are very similar.
The PBE functional may be viewed as slightly favorable because 
it was designed to have a simpler form and derivation, and
already revised versions exist (see, e.g., reference \onlinecite{Hammeretal}).

The nickel basis set from reference \onlinecite{Towler} was employed, with the
innermost $[4s3p1d]$ exponents being unchanged. In addition, 2 $sp$
functions with exponent 0.63 and 0.13 and a $d$-function with exponent
0.38 as optimized in calculations for the bulk were used, so that
the basis set as a whole was of the size $[6s5p2d]$.
The chlorine and potassium 
basis sets were the ones used in previous studies on 
Cl/Cu(111) and K/Cu(111), respectively\cite{DollHarrison2000,Doll2001KCu}.
For the calculations on the free atoms, enhanced basis sets with
additional diffuse functions were used \cite{basisdetail}.
For the numerical integration, the potential was fitted with
auxiliary basis sets (for Cl the same as in reference
\onlinecite{DollHarrison2000}, for K as in reference
\onlinecite{Doll2001KCu}, and for Ni the same as for Cu in reference
\onlinecite{Doll2001KCu}). 
A $\vec k$-point net with 16 $\times$ 16 points in the surface
Brillouin zone and
16 $\times$ 16  $\times$ 16 points for the bulk was used. A 
smearing temperature of 0.001 $E_h$ was applied to the Fermi function
($1 E_h$=27.2114 eV=315773 K). 
This value for the temperature was chosen  
relatively low 
to ensure that the magnetic moment is not artificially modified by a too high
value. For more details and tests on the computational parameters and
basis sets, see also the appendix.

The calculation of the charge and spin population is based on the Mulliken
population, 
which is actually due to McWeeny\cite{McWeeny,Mulliken} 
(see also reference \onlinecite{McWeenyIJQC}). 
The Mulliken charge is defined as
$\sum_{\mu \in A}\sum_{\nu}P_{\mu\nu}S_{\nu\mu}$
(with density matrix $P$, overlap matrix $S$ and $A$ the set of basis 
functions for which the population is computed). In the case of the
spin population
the density matrix for the corresponding spin polarization has to
be used. This type of population analysis thus 
depends on the basis sets. It will be questionable if basis sets
of different quality are used for various atoms: if one basis set
is very small and the other very large, and especially if the larger
basis set also has more diffuse exponents  (i.e. smaller exponents,
which results in a Gaussian with a relatively large spatial extension),
the larger
basis set may describe a part of
the charge of the atom with the smaller basis set as well,
and thus the Mulliken charge will be artificially large for the atom
with the larger basis set. It is thus important that balanced basis
sets are used. 
Concerning the Mulliken populations computed in this article,
the population can be expected to be fairly
reliable in the case when identical basis sets are used for the
various atoms in a slab, so that the magnetic moments computed for
the various layers of
a nickel slab should not have a too large error due to the choice of the
basis set. Also, comparing results for various adsorption sites
should not be too unreasonable, because the basis sets for the systems
are identical. This means, that, for example,
a relative comparison of 
the Mulliken charge for chlorine on the fcc site with
the charge on the top site should not be too unrealistic, although
the absolute value of the Mulliken charge may have a larger error. 
A more difficult 
issue is however, to compare the chlorine charge for Cl/Ni(111)
with the one for Cl/Ag(111) or Cl/Cu(111) because different basis
sets for the metals are used and because of the different geometries.

\begin{figure}
\caption{The threefold hollow structures considered for Cl,
adsorbed on the Ni(111) surface, at a coverage of one third of a monolayer,
$(\protect\sqrt 3 \times \protect\sqrt 3)$R30$^\circ$ unit cell. 
The nickel atoms in the top layer 
are displayed by open circles. The displayed chlorine adsorption sites are
the threefold hollow sites (fcc or hcp hollow,
filled circles). Note that these threefold hollow
sites can not be distinguished in this figure. The ratio of the
radii is chosen
according to the computed values, i.e. using a nickel radius of 1.25 \AA
\ and a chlorine radius of 1.09 \AA. The top site would be vertically
above a nickel atom, the bridge site vertically above a point which is in
the middle between two neighboring nickel atoms.}
\label{geometryfigure}
\centerline
{\psfig
{figure=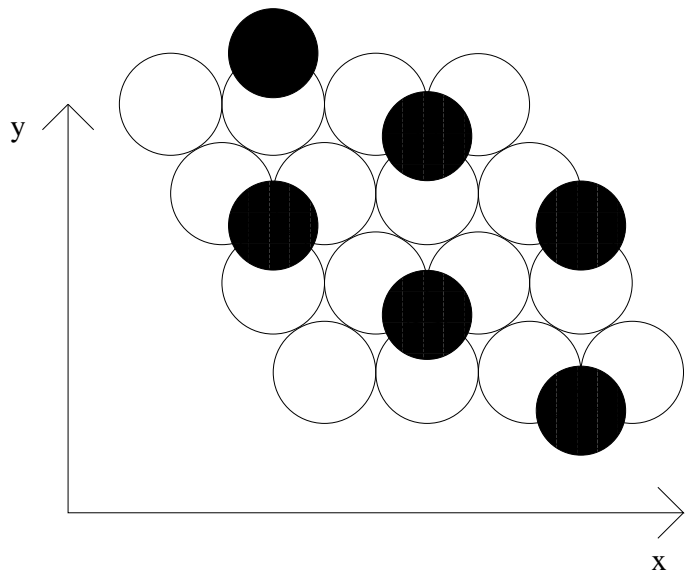,width=15cm,angle=270}}
\end{figure}

\begin{figure}
\caption{Potassium,  
adsorbed on 
the Ni(111) surface, at a coverage of one fourth of a monolayer,
$(2\times 2)$ unit cell. The nickel atoms in the top layer 
are displayed by open circles. The displayed potassium adsorption site is
the top site vertically above the nickel atoms (filled circles).  
The ratio of the radii is chosen
according to the computed values, i.e. using a nickel radius of 1.25 \AA
\ and a potassium radius of 1.54 \AA.}
\label{geometryfigure2}
\centerline
{\psfig
{figure=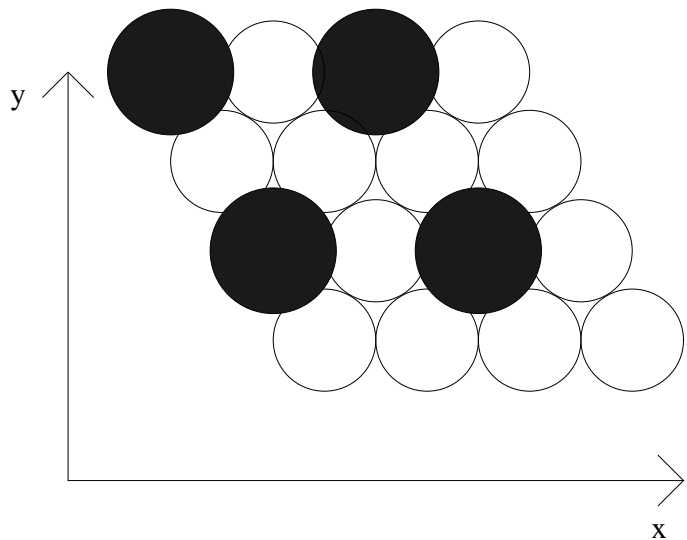,width=15cm,angle=270}}
\end{figure}

The adsorbate systems Ni(111)
$(\protect\sqrt{3} \times \protect\sqrt{3})$R30$^\circ$-Cl
and Ni(111)${(2 \times 2)}$-K 
were modeled with a slab consisting of 3 nickel
layers and the adsorbate layer, with the adsorbate layer on one side
of the slab. The model is truly two-dimensional and not periodically
repeated in the third dimension. This is different from the more common
approach where periodicity in three dimensions is used, and
a number of vacuum layers is required to separate the periodically repeated
slabs. In the latter case, this also introduces an artificial electrical
field due to the periodic boundary conditions in the third dimension,
which is usually compensated for by an additional dipole field
\cite{NeugebauerScheffler}. The approach used by the CRYSTAL code for
slabs avoids this artificial field as there are no periodic boundary
conditions in the third dimension, and thus there is no correction
required, and there is no parameter for the number of vacuum layers either.
The simulation was performed for the 
fcc, hcp, bridge and top site
(see figures \ref{geometryfigure} and \ref{geometryfigure2}). To ensure
the stability of the results with respect to the number of layers, 
additional calculations with thicker
slabs were performed for the systems Cl/Ni(111) (which is easier to
converge than K/Ni(111)).

\subsection{Properties of bulk Ni and clean Ni surfaces}

Results for cohesive properties of bulk Ni  are  displayed in Table
\ref{Nibulktable}. The computed values in this work reproduce
essentially the ones obtained with the gradient corrected functional
of Perdew and Wang, with the plane-wave
pseudopotential method \cite{Moroni}. 
The binding energy is overestimated, compared to the experiment.
Also, the energy splitting between $d^8s^2$ and $d^9s^1$  is not
well reproduced (the $d^9s^1$ state should be lower by
-0.03 eV, after averaging the experimental energies over spin-orbit
components\cite{Moore,MartinHay1989}). 
This is however, not a serious problem, as the
orbital occupancy in the solid is different from the atom and
thus the error cancellation does not work too well, and a density
functional description may be too crude to describe these effects.
The splitting between  $d^8s^2$ and $d^9s^1$ state is a problem
well known in quantum chemistry. The relativistic effect to the
energy splitting between these states was 
computed to be of the order of 0.3 eV 
\cite{MartinHay1989,RaghavachariTrucks1989} (i.e. in a non-relativistic
calculation as the present one, the splitting should be
-0.33 eV instead of -0.03 eV) and thus is not negligible.
With highly accurate
configuration interaction schemes \cite{RaghavachariTrucks1989},
the calculated splitting is -0.18 eV; i.e. 
there is still an error of 0.15 eV, compared to the experimental splitting, 
after relativistic effects were subtracted from the experimental value.
The value computed here
(-0.035 $E_h$ $\hat=$ -0.95 eV); is in reasonable agreement with
other results from literature, e.g. in the range of -1.1 to 
-1.2 eV, both at the level of the local density approximation
(LDA) and the level of the
generalized gradient approximation (GGA), with the PW functional
\cite{Moroni}. This difficulty
in obtaining the correct energy splitting between 
$d^{n-1}s^1$ and $d^{n-2}s^2$
was also identified as a general problem for the transition metals
\cite{HarrisJones}, with the LDA, where a bias towards d$^{n-1}s^1$ of
$\sim$ 1 eV was found. With a different gradient corrected functional
(B-LYP), an energy splitting of -0.82 eV was computed\cite{Russoetal}.
There is, however, no reason why other properties should be affected
by this problem.

The band structure is displayed in
figure \ref{nibulkbandstructure}. It
 agrees well with the early tight binding 
\cite{Langlinais1972,WangCallaway1974} or augmented plane 
wave\cite{Connolly1967} results, and thus supports
the validity of the applied approach, relying
on Gaussian type orbitals.

Surface formation energies (in $E_h$ per surface atom)
of the clean, unrelaxed surfaces are displayed in Table
\ref{Nisurfsummtable}. The surfaces were modeled with a finite number of
Ni layers (typically from 1 to 9).
The PBE energies obtained here are within literature values. For the
(111) surface, the surface energy 
was additionally computed with the Perdew-Wang functional, and is
identical to the one with the PBE functional.

Finally, in Table \ref{surfacemagneticmoment}, the magnetic moments of
the layers of the unrelaxed
slabs are given. In all cases, they were computed using
a 9 layer slab. They are in good agreement with literature values.
For the (111) surface
it is found that the second layer in the slab has the highest magnetic moment
and not the top layer, in agreement with two of the published articles. 
For comparison, for the (111) surface, relaxation of the top layer was 
additionally taken into account. This resulted in a small inwards
contraction by 0.03 \AA, which is consistent with experimental
\cite{Luetal1996} and theoretical \cite{Mittendorfer1999}
findings. The magnetic moments were virtually unchanged,
and thus relaxation does not seem to explain why in one publication
it was observed that the outermost layer should have the highest moment
\cite{Mittendorfer1999}. For comparison, also the Perdew-Wang functional
was used to compute the moments of the unrelaxed (111) surface, and the
results of the PBE functional were essentially reproduced.
As explained earlier, the calculation of the magnetic moments is based
on the Mulliken population analysis. The error should, however, not be
too large as identical basis sets are used for all the nickel atoms
of the slabs.

\section{Cl on the Ni(111) surface}
\label{chlorineadsorptionsection}

As a first adsorbate system, the adsorption of Cl on the Ni(111)
surface was investigated. This system was studied experimentally
with LEED (low energy electron diffraction), Auger electron spectroscopy,
work function measurements and flash desorption \cite{Erley1977},
surface extended X-ray absorption fine
structure (SEXAFS) and soft X-ray standing wave (XSW)
\cite{Funabashi1989,Funabashi1991,Takata1992}, 
 and angle-resolved photoemission fine-structure measurements
(ARPEFS)\cite{Wangetal}. At low coverages there were indications
of a $(2 \times 2)$ pattern, followed by a 
$(\protect\sqrt{3} \times \protect\sqrt{3})$R30$^\circ$ structure, and
a complex $\left( 2 \ 1 \atop {4 \ 7} \right)$ pattern. The 
$(\protect\sqrt{3} \times \protect\sqrt{3})$R30$^\circ$ structure is
the one which has been extensively studied experimentally and thus
also the calculations will focus on  this pattern.
There was agreement that 
the fcc hollow was the preferred site (with an fcc:hcp population
ratio of around 85:15\cite{Takata1992}, and the Cl-Ni bond length was
measured to be 2.40 $\pm$ 0.04 \AA \cite{Funabashi1989}, 2.33 
$\pm$ 0.02 \AA \cite{Funabashi1991} or 2.332 $\pm$ 
0.006 \AA \cite{Wangetal}. This corresponds to an interlayer distance of
1.92 \AA \cite{Funabashi1989}, 1.83  \AA
\cite{Funabashi1991} or 1.837  \cite{Wangetal}.
The first nickel layer was found to be practically
unrelaxed in two experiments\cite{Funabashi1989,Funabashi1991},
whereas in the other article, a 5\% contraction with respect to the bulk 
value was deduced \cite{Wangetal}. 

It is thus an interesting issue to compare with results from the
simulations. These results are presented in Table \ref{ClonNitable}.
A structural optimization was performed with 3 and 4 nickel layers.
An additional single point calculation with 5 nickel layers was
performed for fcc and hcp site, where the geometry from the thinner
slabs was used, with one additional nickel layer at the distance
corresponding to the bulk.
The vertical positions of the 
chlorine adatoms and of the first nickel
layer were optimized. Lateral relaxations are not possible for fcc
and hcp site. As bridge and top site are much higher in energy, 
lateral relaxations were not taken into account for these sites either.
Comparing the data for three and four layers, we see that the
geometry is well converged, and the difference in the
energy splitting between the
sites is within what is expected as the numerical noise (e.g.
fcc vs. hcp site, the energy splitting is 0.7 $mE_h$ with three nickel 
layers,
0.2 $mE_h$ with four layers, and 0.6 $mE_h$ with five layers). 
Firstly, we note that the fcc hollow is indeed found to be lowest in 
energy, with the hcp hollow being nearly degenerate at an energy
difference of only 0.2-0.7 $mE_h$ ($\sim$ 60-220 K). 
Although the numerical noise is not
negligible when energy differences on this small scale are computed
(see appendix), this energy difference would be compatible with the
simple thermodynamic estimate given in reference \onlinecite{Kadodwala},
with copper as substrate.
The bridge site is higher in energy
by $\sim$ 3 $mE_h$ (or more than 900 K), 
the top site by 18 $mE_h$, and thus thermodynamically fcc and hcp sites
are clearly favored. The energy splitting between fcc (or hcp site), and
the bridge site, may also be a crude estimate for the diffusion barrier
from one threefold-hollow site to another one.
These findings are
actually very similar to those 
\cite{DollHarrison2000,DollHarrison2001,Jia2003} for 
Cu(111)$(\protect\sqrt{3} \times \protect\sqrt{3})$R30$^\circ$-Cl and 
Ag(111)$(\protect\sqrt{3} \times \protect\sqrt{3})$R30$^\circ$-Cl,
and also for the system Ag(111)$(1 \times 1)$-O \cite{Lietal2002}:
on the (111) surface,
the highly coordinated sites are clearly lowest in energy for adsorbates
with a small radius.
The bond lengths from chlorine to the nearest nickel
$d_{\rm {Cl-Ni \ nn}}$
also increase with increasing coordination number
and  the binding energy increases with increasing coordination number,
as is to be expected from Pauling's argument \cite{Pauling}:
one strong Cl-Ni bond in the case of the top site will be shorter 
than the three weaker bonds in the case of fcc and hcp, but the
three bonds as a whole are energetically more favorable.
The computed bond length agrees well with the values from the literature
\cite{Funabashi1989,Funabashi1991,Wangetal}.

There are two different experimental results for the interlayer distance
between first and second nickel layer. The computed value from
this article agrees very well with the experimental from references
\onlinecite{Funabashi1989,Funabashi1991}, 
whereas the large contraction
of 0.10 \AA \ \cite{Wangetal}
with respect to the bulk appears to be unlikely.
A smaller contraction would also be in line with simulations
with Cu or Ag as substrate \cite{DollHarrison2000,DollHarrison2001,Jia2003}.

An effective radius for Cl on Ni(111) could be computed by subtracting
an effective nickel radius of $\frac{3.53}{2\sqrt{2}}$\AA \ from
the bond length $d_{\rm {Cl-Ni \ nn}}$ of 2.34 \AA. This would result
in a radius of 1.09 \AA, in agreement with the computed values
\cite{DollHarrison2000,DollHarrison2001}
for Cl on Cu(111) (1.12 \AA) or Cl on Ag(111) (1.17 \AA). Thus the radius for
Cl is intermediate between the atomic (0.99 \AA) and ionic (Cl$^-$; 1.81 \AA)
\cite{Kittel}. 
The bond length $d_{\rm {Cl-Ni \ nn}}$
is also in agreement with
experimental values for systems with similar atomic radii for
the substrate (e.g., bond lengths of
2.39 \AA \ for the substrate Cu(111)\cite{Crapper1986},
2.38 \AA \ for Rh(111)\cite{Shard1999}, 2.39 \AA \
for Pd(111)\cite{Shard2000}). 

In Table \ref{Clpopulationtable}, the Cl population
is displayed. The data is extracted from the calculations where
three nickel layers were used to model the substrate, but it 
is virtually identical to the data obtained with four nickel layers.
The negative charge is largest for the top site, smaller
for the bridge site and smallest for hcp and fcc site. This is
consistent with the position of the 3$s$ core eigenvalue relative to
the Fermi level, which is lowest
for hcp and fcc, increasing for bridge and largest for the top site
(because the negative charge of the chlorine adsorbate is increasing and thus
the core levels are destabilized). Note that the core level for
majority spin is also slightly lower (by $\sim$ 0.002 $E_h$) than
for minority spin. 
The density of states is visualized in figure \ref{ClNiDOS},
projected on the chlorine basis functions, and projected on all
basis functions (i.e. the total density of states). The different
occupancy of the two spin states becomes visible. Note that the
chlorine spin is mainly due to the different population of the second
and third $p$-shell, and not of the fourth (if we order the shells
according to the size of the exponents, starting with the highest).
This means that the outermost and thus most diffuse exponent does not
carry a huge spin and thus the chlorine spin is not an artifact
of the Mulliken population (if the spin was on the most diffuse
function only, one might argue that this basis function also describes
the spin on the nickel atoms and thus the Mulliken 
population analysis would be questionable).

The computed Mulliken charge 
is also in agreement with
the experimental estimates of 0.1 to 0.2 $|e|$ derived from work
function measurements\cite{Erley1977}. The overlap population
$\sum_{\mu \in A}\sum_{\nu \in B}P_{\mu\nu}S_{\nu\mu}$
(with density matrix $P$, overlap matrix $S$ and $A, \ B$ the set of basis 
functions of atoms $A$ and $B$ for which the overlap population is computed)
is 0.09 for Cl and the nearest nickel atoms. 
This may indicate that
there is no evidence for a strong covalent contribution to the binding
mechanism, similar to non-magnetic substrates
\cite{DollHarrison2000,DollHarrison2001,Doll2001KCu,Doll2002},
and now
also in the case of a 
magnetic substrate. 
For comparison, the overlap population between two
nearest nickel atoms in the top nickel layer is 0.10.
However, this data should not be over-interpreted, and the 
overlap population should rather be viewed as a rough qualitative
measure and not a quantitative measure for covalency.
Another indication that the covalency is not strong might be that
in the case of the bridge site, where $p_x$ and $p_y$ orbitals do not have
the same occupancy, the orbital with the larger overlap with
the neighboring nickel atoms ($p_x$ in our choice of geometry) has
a slightly smaller population than $p_y$
(the total Cl $p$ population is
3.73 for $p_x$, versus 3.82 for $p_y$).

Charge density, charge density difference 
and spin density are visualized in figure 
\ref{ClonNichargespindensity}. As usual, 
the charge density difference is more
informative as the charge density. 
The transfer from the top nickel layer to the chlorine atom
becomes obvious. In contrast to the charge density, the spin density
also offers more information:
it is very interesting that the spin density
varies relatively strong still 
for the atoms in the third layer. This is
confirmed by the data from Mulliken population analysis for the
spin (table \ref{Clspintable}). The magnetic moments depend relatively
strong on the number of layers, and the analysis has been performed
by extracting the data from a single-point calculation for a
slab with 5 nickel layers, with an identical geometry as for
the 3-layer slab, i.e. the additional layers were added at a distance
as in the bulk. 

The data is presented in such a way that one number is given, if
all the atoms in a layer have the same distance to the chlorine atom, and
two numbers, if there are two different distances for the atoms in one
layer possible. 
For example, in the case of the fcc site, looking at  
the atoms in the first or second layer, 
all three nickel atoms within a supercell
have the same distance to the adsorbed chlorine. However, in the third 
layer there will be one atom which is closer than the other two. The
population of this closer atom (0.59) is displayed in the left column of
the results for the $3^{rd}$ layer, the data for the other two atoms (0.64)
in the right column.

It is interesting to note that those nickel atoms in the first layer
which are closer to the chlorine atom have a reduced magnetic moment. This
is strongest in the case of the top site where the atom vertically under
the chlorine has its population reduced to 0.55, versus a moment of 0.76
for the other two atoms. It is also apparent for the bridge site
(0.66 vs 0.73), and in the case of fcc and hcp site the data is identical
because the nickel atoms in the first layer are identical by symmetry.

In the second layer, the spin population is higher for those atoms
which are closer to the adsorbate (this is possible for the hcp and
bridge structure). In the third layer, again those atoms which have
a shorter distance to the adsorbate, have there moment reduced.
These oscillations are still visible in the fourth and fifth layer.

Although this data depends relatively strong on the number of layers,
it becomes apparent that there are oscillations in the spin
density. It is also interesting to compare with the total charge:
the latter quantity is virtually constant within the layers,
from the second layer on, and has a value identical to that of
a clean slab from the third layer on (also table \ref{Clspintable}).
Only in the first layer, the atom(s) closer  to the chlorine adsorbate
have a slightly higher charge (i.e. more electronic charge) than the
atom(s) which are further apart (visible for bridge and top site).

Finally, the chlorine adsorbate has a magnetic moment of $\le$ 0.1 for
all sites (the smallest value for the top site). The moment is always
parallel to the nickel spin.

It is interesting to note that in the simulations
for CO on Ni(110) \cite{Geetal2000}, it was also found
that the moment of the atom(s) closest to the CO was reduced.

\section{K on the Ni(111) surface}
\label{potassiumadsorptionsection}

In addition, the system Ni(111)${(2 \times 2)}$-K was studied with
this approach. Experimentally, 
in a LEED study at various coverages, this was found
to be the only commensurate structure \cite{Chandavarkar1988}.
In  a further LEED study\cite{Fisheretal,Kaukasoinaetal1993},
the top site was found to be occupied, with a K-Ni distance of
2.82$\pm$ 0.04 \AA, a vertical rumpling of the first Ni layer of 0.12 \AA,
and a lateral displacement of 0.06 \AA \ of the three nickel atoms
in the top layer which are not vertically under the K adsorbate.
The adsorption site was later confirmed in SEXAFS 
measurements\cite{Adler1993},
and a larger bond length of 2.92$\pm$ 0.02 \AA \ was obtained. 
With the same technique, also the bond length for the system K/Cu(111)
was computed and deduced to be by 0.06 $\pm$ 0.04 \AA \ larger.
From 
ARPEFS\cite{Huangetal1993},
a top site adsorption with K-Ni bond-length of 3.02 $\pm$ 0.01 \AA \
was obtained.
In addition, the first to second layer spacing was reduced by 0.13 \AA \
to 1.90 $\pm$ 0.04 \AA,
but no substrate rumpling in the top layer was observed. Lateral
displacements were not found either (0.00 $\pm$ 0.09 \AA) in this
experiment.
Finally, photoelectron diffraction (PhD) \cite{Davis1994} resulted
in a bond length of 2.87 $\pm$ 0.06 \AA, and a reduced first
to second layer distance of 1.86 $\pm$ 0.06 \AA \ was observed, with
virtually no rumpling in the first layer (0.01 $\pm$ 0.09 \AA).

The experiments thus consistently support the top site as the 
adsorption site, but the geometrical parameters are not identical.
Moreover, a comparison with the system K/Cu(111) is very interesting,
so that the influence of the partially occupied $d$-bands can be 
investigated.

The results of the simulations are summarized in Table \ref{KonNiEnergie},
with geometrical parameters as defined in figure \ref{geometricalparameters}.
Initially, the potassium position and the position of the top nickel layer was
optimized (i.e. the uniform relaxation of the top nickel layer). In
a second step, a vertical rumpling of the nickel atoms in the top layer 
was allowed, as this is expected to be important for the top site
occupancy. A further refinement would have been the additional possibility
of lateral relaxations. These lateral relaxations were, 
however, found to be of minor
importance for the related system K/Ag(111), where the energy
was lowered by the order of typically 
0.1 - 0.2 $mE_h$, when these relaxations were
allowed\cite{Doll2002}. 
Thus, lateral relaxations were taken into account additionally
only for 
the top site, which was found to be lowest in energy, and in order
to compare with the experimental data for the lateral displacement.

Firstly, the top site is found to be the most favorable site. This is
in agreement with the experimental
findings \cite{Fisheretal,Kaukasoinaetal1993,Adler1993}. 
By disabling
the possibility of substrate rumpling, it is also demonstrated that
substrate rumpling is crucial for top site occupancy, as it lowers
the energy by 1.6 $mE_h$, but only 0.4 $mE_h$ for the other
sites. These findings are actually similar to the
system K/Cu(111) \cite{Doll2001KCu}. The top site thus becomes the most
stable site because the nickel atom underneath is pushed into the substrate
relative to its neighbors and the potassium with its relatively large
radius thus also overlaps more with the second nearest nickel neighbors.

The computed bond length 
(2.79 \AA) is also very similar to K/Cu(111) 
(2.83 \AA) \cite{Doll2001KCu} so that these calculations do not
indicate a major difference compared to Cu.
Experimentally, using the same technique for both adsorbate systems, 
it was measured to be slightly shorter
and it was speculated that this was due to greater adatom-substrate
interactions involving the partially filled Ni $d$ bands\cite{Adler1993}. 
The possibility of a lateral displacement of the nickel
atoms as observed experimentally (0.06 \AA)\cite{Fisheretal}
was investigated. However, this could not be confirmed and
the energy was found to be lowest for the case where this
lateral displacement is exactly zero.

Bridge, fcc and hcp site are found to be energetically degenerate.
The fact that all the sites are energetically very close
(e.g. much closer than in the case of chlorine)
is consistent with earlier findings 
\cite{NeugebauerScheffler,Doll2001KCu,Doll2002}
and was
explained due to the large radius of the adsorbate which makes
it experience only a small  substrate electron density corrugation
\cite{NeugebauerScheffler}. This small energy splitting between the
various sites is also consistent with the experiment
because of various observations\cite{Adler1993}: firstly,
the SEXAFS data indicated enhanced
in-plane motion resulting from a wide and shallow potential well.
Secondly, no SEXAFS signal was observed for low coverages below 0.13 
monolayers. It was thus argued that lateral motion was essentially hindered
by the presence of other adatoms. Finally, with increasing temperature
(already in the range of 120-145 K)
the SEXAFS amplitude was found to decrease. This was explained due
to the occupancy of other sites, which leads to a destructive interference.

The bond length $d_{\rm {K-Ni \ nn}}$
increases with the number of nearest nickel neighbors in
the order top, bridge, fcc and hcp. The effective radius of the
potassium adsorbate is 2.79-$\frac{3.53}{2\sqrt{2}}$ \AA=1.54 \AA \ for
the top site. This
number is in good agreement with experimental values for potassium
when occupying the top site \cite{DiehlMcGrath}, or the computed
value for K/Cu(111) (1.55 \AA\cite{Doll2001KCu}).

Looking at the populations in Table
 \ref{Clpopulationtable},
we see that they ($\sim$ 0.3 $|e|$)
are very similar for all considered
sites, only for the top site the charge is slightly less positive.
In agreement with this, the core eigenvalues of the $3s$ and $3p$
orbital are virtually independent of the site. The core levels
are identical for both spin polarizations which agrees with the
finding that the magnetic moment of the potassium adatoms is negligible
(see below). In the calculations for K/Ag(111) \cite{Doll2002}, it
was recently shown that the adsorbate charge and thus also the 
core eigenvalue  (and other properties such as the bond
length) change with the coverage. 
Such a change in the core eigenvalues was observed experimentally for 
alkali metals on metallic surfaces, e.g.
for the systems Na/Cu(111) and Na/Ni(111) \cite{Shietal1993}, 
or for alkali metals
on Ru(100) \cite{Sheketal1990} or W(110) \cite{Riffeetal1990}.

The overlap population
is $\sim$ 0.03 for
K and the nearest nickel atom, so that there is also in this case
no evidence for a strong covalent contribution to the binding mechanism.

The density of states is displayed in figure \ref{KNiDOS}. 
The integrated density of states, when projected on the potassium
basis functions, differs so little for majority and
minority spin that there is nearly no magnetic
moment. For the top site, the spin is even antiparallel to the nickel sites,
in contrast to the other sites (and in contrast to chlorine).

Again, the charge density is not very informative, and
it is more interesting to look at 
the charge density difference (figure \ref{KonNispin}). Electrons
have flown from the adsorbate mainly to the top nickel layer
(and a relatively large part
to the atom vertically under the potassium adatom). Thus, the
charge of the potassium overlayer has decreased.
It is apparent that the spin density of the nickel atom vertically
under the top atom is slightly lower than the spin density of its neighbors
in the same layer, as the contours are closer to the nucleus. This is
confirmed in the Mulliken spin population (table \ref{Kspintable}).
Again, those atoms in the top layer which are closest to the potassium
adsorbate, have their moment reduced.
No noteworthy variation is found in the second and third nickel layer,
and the values for the spin  population are very close to
the ones for the clean Ni slab.
The moment of the potassium adsorbate is negligible, as its magnitude
is $\le$ 0.003.
Also, the charge only varies in the first layer, and already the second
and third layer have charges which are identical to that of a clean slab.

There are thus no oscillations of the spin population beyond the first 
nickel layer, probably because the potassium magnetic moment is much
smaller than that of chlorine.

\section{Summary}

We have studied Ni bulk, the clean low index surfaces and 
the adsorption of chlorine and potassium on the Ni(111) surface.

The results for the adsorption  geometry confirm the
experimental findings. It turns out that there is no huge
difference to copper as substrate; the preferred adsorption site and
energies are very similar. 

The reason for the preference
of the top site for K is substrate rumpling which helps to increase the
overlap of the K adsorbate with the second nearest nickel neighbors. 
This rumpling is thus crucial
for the site preference. In the case of chlorine, the 
threefold hollow sites are very close in energy. In general, the adsorbate
atoms
prefer sites with the highest possible coordination number, or they try
to increase the overlap to neighboring atoms by substrate rumpling.

There appear to be no strong covalent contributions to the binding so
that the mechanism appears to be mainly an ionic bond (although
the charge transfer is rather small), and in the case of 
potassium, additionally a metallic binding mechanism. However,
the quantitative analysis of the binding mechanism is certainly difficult.
The computed charges seem to be fairly reliable, but the overlap
population is probably only a qualitative tool to estimate the degree
of covalency.

Charges and core levels were found to be consistent, i.e. with
increasing negative charge of the adsorbate, the core levels of the
adsorbate are slightly destabilized.
It is also
demonstrated that the magnetic moments computed with the help
of the Mulliken population analysis agree well with data from other
schemes, e.g. pseudopotential plane wave. The calculation of work
functions, however, seems to be difficult with a local basis set
and the results are strongly basis-set dependent.

Finally, the magnetic moment of those nickel atoms closest to the
adsorbate are always reduced. Oscillations of the spin population
within one nickel layer are notable for the first three layers for Cl/Ni(111),
but only for the first nickel layer for K/Ni(111), probably because
of the very small magnetic moment of the potassium adsorbate, 
compared to chlorine.

\section{acknowledgments}

All the calculations were performed at the computer centre of
the TU Braunschweig (Compaq ES 45).

\appendix
\section{Test of computational parameters}
The calculations require various sets of parameters, with the
most important being: a finite number
of $\vec k$-points, a finite temperature to smoothen the integrand
and thus to facilitate the numerical integration of the exchange-correlation
functional, basis set parameters, and truncation schemes to reduce the
number of matrix elements (i.e. integrals of operators such as 
kinetic energy, nuclear attraction, coulomb and exchange, with the
basis functions). In this first part of the
appendix, the stability of the results
with respect to the variation of parameters concerning the grid
and the parameters concerning the selection of the integrals is tested.
For further tests, see also the extensive tests performed for lithium
\cite{KlausNicVic}, or tests for copper\cite{DollHarrison2000} 
and silver\cite{Doll2002} ($\vec k$-points, smearing temperature).
In tables \ref{ClNiparametertest} and \ref{KNiparametertest}, 
the binding energies are computed with the default
grid and truncation parameters, and with an improved grid and
two sets of stricter and thus more accurate truncation parameters.

The improved grid has roughly six times more sampling points than the default
grid for the various structures.
As is displayed in the tables, the grid has virtually no impact on the
energy splitting. For Cl/Ni(111), the binding energy changes by
at most 0.2 $mE_h$ for fcc, hcp and bridge site, and only for
the top site an energy change of 0.9 $mE_h$ is observed, with respect
to the default grid. This has, however,
no impact as the top site is much higher in energy anyway.
For K/Ni(111) the absolute value of the
binding energy is shifted by
0.7 $mE_h$, and the relative energies of the various sites
are identical.

There are two truncation parameters in calculations when
no Fock exchange is involved\cite{Manual}: 
a general one ("ITOL 1") which leads to
a neglect of all integrals
with an overlap below a certain threshold (default: 10$^{-6}$), and
a second parameter  ("ITOL 2")
which leads to the evaluation of the coulomb integrals at a
lower level of accuracy by means of a multipolar expansion,
if the overlap is below another threshold (default: 10$^{-6}$). As
the results turned out to be more sensitive to these parameters, two sets of
higher thresholds were used (10$^{-7}$ for ITOL 1 and 2, and 10$^{-8}$
for ITOL 1 and 2). 
In the case of chlorine, the largest error is found for the
splitting between fcc and hcp site: it ranges from 0.7 to 2.1 $mE_h$,
for the various parameters, with the fcc site always being more
favorable. The bridge site is always clearly
higher than the fcc site by 3.3 to 4.2 $mE_h$, and the top site
is much higher in energy.
For K/Ni(111), the impact of these parameters is smaller, and
for example, the splitting from the top site to the site closest
in energy ranges from 0.5 $mE_h$ to 0.8 $mE_h$. 

As a whole, the impact of these computational parameters is certainly
not negligible, but the conclusions are not affected. The fcc site
is favored for Cl/Ni(111), with the hcp site being nearly degenerate.
In the case of K/Ni(111), the top site is favored, with the other
sites being very close in energy.

\section{Basis set dependence of the results}
\label{Basissetappendix}

The choice of the basis set is of high importance for the quality
of the results. In this part of the appendix, the
results of various tests are displayed, to investigate the dependence
of the results on the basis set. These tests focus on nickel bulk,
the clean surfaces, and Cl/Ni(111).

Firstly, one might consider reoptimizing the exponents for the
clean surface, for example keeping the exponents of the inner layer 
like in the bulk
and reoptimizing the exponents of the outermost layers. 
This would result (with a slab with three layers)
in exponents of 0.12 ($sp$, instead of 0.13 for the
bulk) and 0.38 ($d$, as in bulk nickel) for the outer layers. 
This is thus a tiny change only
and the results are virtually identical, for example the surface energy
per atom changes by 0.6 $mE_h$ which is negligible. If an outermost
$sp$-exponent of 0.12 instead of 0.13 is used for the bulk, 
exactly the same results as in table \ref{Nibulktable} are obtained.

There is however a property which depends extremely strong on the
basis set: the computed work function changes drastically
when the outermost exponent is changed by a tiny value. For example,
with the $sp$-exponent used for all the calculations (0.13), 
the work function of the Ni(100) surface is 0.142 $E_h$ 
(exp.\cite{Baker,Michaelson}: 0.192 $E_h$),
of the Ni(110) surface 0.129 $E_h$ (exp.\cite{Baker,Michaelson}: 
0.185 $E_h$), and of the
Ni(111) surface 0.155 $E_h$ (exp.\cite{Baker,Michaelson}: 0.197 $E_h$). 
This data
is practically stable with respect to the number of layers (i.e. 
3 layers are sufficient to obtain a stable number).
When, 
however, for example, the outermost $sp$-exponent
of the outer layers is changed from 0.13 to 0.12, the
work function of the Ni(111) surface changes by 0.01 $E_h$ to 0.165 $E_h$,
and it further increases with smaller exponents (e.g. 0.183 $E_h$, if
the outermost $sp$-exponent of the outer layers is 0.10 instead of 0.13).
Also, increasing this exponent leads to a lower work function. 
The same problem shows also up with other metals, such as for
example copper or silver (and also when the exponents of all the
layers and not just the outermost layers are changed).
It appears thus that the work function is a quantity which is not
well described with a local basis set. 
The work functions may be qualitatively reasonable, but a quantitative
comparison seems unadvised. Thus, also the work function data computed
for Ag and the adsorption thereon
\cite{Doll2002} has to be considered with caution. A problem with
computing work functions with a local basis set was already discussed
earlier \cite{Boettgeretal1995}.

The results thus indicate that 
the only property which appears to be strongly basis set dependent is
the work function
(total energy, surface energy, 
bulk modulus and lattice constant are essentially
stable with respect to small variations in the basis set). 
Mulliken charge and magnetic moment are stable if an identical basis
is used for all the nickel atoms in a slab.
Of course, if
basis sets of different type are used for the inner layer of
a slab and the outermost layers, the Mulliken population also gets
less reliable.

Further tests were performed  for the adsorbate system Cl/Ni(111). 
One test was done by choosing a more tight outermost
$sp$-exponent of 0.15 instead of 0.09 for chlorine. When a structural
optimization is performed, virtually the same structural data as in table
\ref{ClonNitable} is obtained (the maximum deviation is 0.01 \AA).
The binding energy is lower by $3 mE_h$ 
because of the poorer chlorine basis set.
However, what is important is that the energy splitting for the various
sites remains practically identical (the binding energy is
-0.1326 $E_h$ for the fcc site,
-0.1320 $E_h$ for the hcp site, -0.1292 $E_h$ for the bridge site,
and -0.1151 $E_h$ for the top site).
Also, properties such as the population, magnetic moment
and the relative position of
the core levels with respect to the Fermi energy are essentially stable
with respect to this change of the basis set
(but again the work function changes strongly).

As a whole, the basis set needs of course to be carefully chosen.
When this is done, the results for the geometry, energies, core
eigenvalues, populations, and magnetic moments are reliable and
not too strongly basis set dependent. The accurate determination of the
work function, however, appears to be difficult with a local basis set.

\onecolumn

\newpage
\begin{table}
\begin{center}
\caption{The ground state properties of bulk Ni. }
\label{Nibulktable}
\vspace{5mm}
\begin{tabular}{ccccccc}
 & &  &  & \\
 & lattice constant $a_0 \ [{\rm \AA}]$ & $E_{coh} \ [E_h] $  & $B$ [GPa] & magnetic moment 
[$\mu_B$] \\
PBE, this work & 3.53 & 0.218$^a$ 0.184$^b$ & 203 & 0.62 \\
PW, this work & 3.53 & 0.220$^a$ 0.185$^b$ & 203 & 0.61\\
Ref. \onlinecite{Moroni}, LDA & 3.43 &   & 255 & 0.59 \\
Ref. \onlinecite{Moroni}, GGA & 3.53 &   & 195 & 0.61 \\
exp. & 3.52 \cite{Gschneider}  &  0.163$^a$, 0.164$^b$
\cite{Gschneider} & 190 \cite{Gschneider} & 0.61 \cite{Kittel} \\	
\end{tabular}
$^a$ energy with respect to a free nickel atom in its $d^8s^2$ state \hfill \\
$^b$ energy with respect to a free nickel atom in its $d^9s^1$ state \hfill \\
\end{center}
\end{table}

%\newpage
\begin{table}
\begin{center}
\caption{\label{Nisurfsummtable}The surface energy $[\frac{E_h}
{\rm surface \ atom}]$ of
the low index nickel surfaces.}
\begin{tabular}{ccccccc}
surface &  PBE, this work  & Ref. \onlinecite{Mittendorfer1999} & Ref.
\onlinecite{Alden} & Ref. \onlinecite{Vitosetal} \\
(100) & 0.038 & 0.031 & 0.039 & 0.036 \\
(110) & 0.056 & 0.046 & - & 0.049 \\
(111) & \hspace{1.7cm} 0.028 (PW: 0.028) & 0.024 & 0.033 & 0.026 \\
\end{tabular}
\end{center}
\end{table}

%\newpage
\begin{table}
\begin{center}
\caption{{Magnetic moments [$\mu_B$] of
the low index nickel surfaces, computed using a slab with 9 Ni layers.}}
\label{surfacemagneticmoment}
\begin{tabular}{cccccccc}
layer &  PBE, this work  & PW, this work & 
Ref. \onlinecite{Mittendorfer1999} (9 layers, \\
      &                  &  & 3 surface layers on both & Ref. \onlinecite{Alden} & Ref. \onlinecite{FuFreeman} & Ref. \onlinecite{Wimmer1984} \\
      &                  & & sides are allowed to relax) &  &  (7 layers) &  (7 layers)\\
(100) surface \\
S & 0.729 & & 0.76 & 0.69 & & 0.68\\
S-1 & 0.634 & & 0.68 & 0.64 & &  0.60\\
S-2 & 0.632 & & 0.66 & 0.66 & & 0.59\\
S-3 & 0.611 & & & 0.64 & & 0.56\\
S-4 & 0.619 &     &\\
(110) surface \\
S   & 0.757 & & 0.76 \\ 
S-1 & 0.637 & & 0.66\\
S-2 & 0.618 & & 0.64\\
S-3 & 0.620 & \\
S-4 & 0.607 & \\
(111) surface \\
S & 0.648 (relaxed: 0.649) & 0.638 & 0.68 & 0.62 & 0.63\\
S-1 & 0.658 (relaxed: 0.657) & 0.649 & 0.65 & 0.67 & 0.64\\
S-2 & 0.622 (relaxed: 0.622) & 0.613 & 0.62 & 0.65 & 0.58\\
S-3 & 0.617 (relaxed: 0.617) & 0.604 &  & 0.63 & 0.58\\
S-4 & 0.613 (relaxed: 0.613) & 0.605 \\
\end{tabular}
\end{center}
\end{table}

%\newpage
\begin{table}
\begin{center}
\caption{Adsorption of Cl on the Ni(111) surface.
$d_{\rm {Cl-Ni \ top \ layer}}$ is the interlayer distance between the Cl 
layer and the top Ni layer, $d_{\rm {Cl-Ni \ nn}}$ is the bond length between
Cl and nearest neighbor Ni. $d_{Ni1-Ni2}$ is the distance between
first and second nickel layer. The distance between second and
third nickel layer $d_{\rm Ni2-Ni3}$ 
is held fixed at the bulk value (and  for the thicker slabs also
$d_{Ni3-Ni4}$ and $d_{Ni4-Ni5}$ are fixed at the bulk value). 
The adsorption energy is the difference 
$E_{\rm {Cl \ at \ Ni(111)}}-{E_{\rm Ni(111)}-E_{\rm Cl}}$.}
\label{ClonNitable}
\begin{tabular}{ccccc}
Site & $d_{\rm {Cl-Ni \mbox{ }top\mbox{ } layer}}$  & 
$d_{Ni 1-Ni 2}$ &
$d_{\rm {Cl-Ni \ nn}}$ & $E_{adsorption}$\\
& [\AA] & [\AA] & [\AA] & $\left[\frac{E_h}{Cl \ atom}\right]$\\
\multicolumn{5}{c}{3 nickel layers} \\
fcc & 1.84  & 2.02   & 2.34 & -0.1358 \\
hcp & 1.85 & 2.02 & 2.35 & -0.1351 \\ 
bridge& 1.90 & 2.02 & 2.27 & -0.1325 \\
top &  2.14 & 2.02 & 2.14 & -0.1177 \\
\multicolumn{5}{c}{4 nickel layers} \\
fcc & 1.83 &   2.02   & 2.33 & -0.1338 \\
hcp & 1.84 & 2.02 & 2.34 & -0.1336 \\
bridge & 1.89 & 2.02 & 2.27 & -0.1307 \\
top & 2.14 & 2.02 & 2.14 & -0.1159 \\
\multicolumn{5}{c}{5 nickel layers} \\
fcc (single point calculation) & & &  & -0.1337 \\
hcp (single point calculation) & & & & -0.1331 \\
\\
exp. (ARPEFS) \cite{Wangetal} & 1.837 &  1.926   & 2.332  &  \\
exp. (SEXAFS, XSW)
\cite{Funabashi1989,Funabashi1991}  & 1.92, 1.83 & 2.03, 2.02
 & 2.40, 2.33  &  \\
\end{tabular}
\end{center}
\end{table}

%\newpage
\begin{table}
\begin{center}
\caption{Charge and position of the $3s$ eigenvalue
for Cl and the $3s$ and $3p$ eigenvalues for K on different adsorption sites,
relative to the Fermi energy. For chlorine, the peaks are at slightly 
different positions for majority and minority bands, for potassium they
are at the same position.}
\label{Clpopulationtable}
\begin{tabular}{cccc}
site & charge, in $|e|$ & $3s$ level, relative to $E_F$ $[E_h]$ & $3p$ level,
relative to $E_F$ $[E_h]$ \\
\multicolumn{3}{c}{Cl on Ni:}\\
fcc    & -0.050 & -0.589 (majority spin); -0.586 (minority spin)  \\
hcp    & -0.046 & -0.586 ; -0.584  \\
bridge & -0.068 & -0.580 ; -0.578\\
top    & -0.152 & -0.537 ; -0.535 \\
\multicolumn{3}{c}{K on Ni:}\\
fcc    & +0.294 & -1.178 (both spins) & -0.583 (both spins) \\
hcp    & +0.295 & -1.178 & -0.582 \\
bridge & +0.291 & -1.178 & -0.583  \\
top    & +0.269 & -1.179 & -0.583  \\
\end{tabular}
\end{center}
\end{table}

\begin{table}
\begin{center}
\caption{Mulliken spin and charge population for Cl/Ni(111), 
on different adsorption sites, extracted from calculations with 5 nickel
layers. The population for the nickel atoms in
the individual layers is given. The number in parenthesis gives the
number of atoms which are identical because of symmetry and thus have an
identical population. The left column corresponds to that atom(s)
in the layer which is closer to the chlorine adsorbate. For example,
in the fcc case, all three nickel atoms of a supercell
in the first and second layer have the same distance
and thus the population is identical,
but one atom in the third layer is closer than the other two atoms
resulting in two different values for the spin population. In the case
of the clean slab, all atoms within one layer are degenerate and thus
only one number is displayed.}
\label{Clspintable}
\begin{tabular}{ccccccc}
site &  1$^{st}$ layer  & 2$^{nd}$ layer  & 3$^{rd}$ layer  &  4$^{th}$ layer
& 5$^{th}$ layer  & Cl   \\ \\
\multicolumn{7}{c}{spin}\\
fcc    & 0.67 (3 atoms) & 0.64 (3 atoms) & 0.59 (1 atom) 0.64 (2 atoms)
& 0.66 (3 atoms) & 0.65 (3 atoms) & 0.08 \\
hcp    & 0.68 (3 atoms) & 0.68 (1) 0.63 (2) & 0.62 (3) & 0.66 (3) & 0.63 (1) 0.66 (2) & 0.08 \\
bridge & 0.66 (2) 0.73 (1) & 0.67 (1) 0.63 (2) & 0.60 (1) 0.63 (2) & 0.66 (2) 0.65 (1) & 0.64 (1) 0.66 (2) & 0.08\\
top   & 0.55 (1) 0.76 (2) & 0.64 (3) & 0.62 (3) & 0.67 (1) 0.64 (2) & 0.65 (3) & 0.04 \\
clean slab  &   0.66 & 0.66 & 0.63 & 0.66 &  0.66\\ \\         
\multicolumn{7}{c}{charge}\\
fcc & 27.97 (3) & 28.02 (3) & 28.00 (1) 28.00 (2) & 28.05 (3)& 27.95 (3) &
17.05 \\
hcp & 27.97 (3) & 28.02 (1) 28.00 (2) & 28.00  (3) & 28.05  (3) & 27.95 (1) 27.95 (2) &
17.05 \\
bridge & 28.00 (2) 27.90 (1) & 28.02 (1) 28.01 (2) & 28.00 (1) 28.00 (2) & 28.05 (2) 28.05 (1) & 27.95 (1) 27.95 (2) & 17.07 \\
top & 28.12 (1) 27.84 (2) & 28.02 (3) & 28.00 (3) & 28.05 (1) 28.05 (2) & 27.95 (3) &
17.15 \\
clean slab & 27.95 & 28.05 & 28.00 & 28.05 & 27.95 \\
\end{tabular}
\end{center}
\end{table}

%\newpage
\begin{table}
\begin{center}
\caption
{Adsorption of K on the Ni(111) surface. 
$d_{\rm {K-Ni1a}}$ is the interlayer
 distance between the K  layer and the layer made of those
 nickel atoms in the first layer closest
to the K layer. $d_{\rm {Ni1b-Ni2}}$ is the distance between the layer made
of those nickel atoms of the first layer which have moved away from
the adsorbate and the second Ni layer. 
$\delta=d_{Ni1a-Ni1b}$ is the rumpling within the first layer.
$d_{\rm {K-Ni \ nn}}$ is the bond length between
K and nearest neighbor Ni. Again, the distance between second and
third nickel layer $d_{\rm {Ni2-Ni3}}$ is held fixed at the bulk value.
The adsorption energy is the difference 
$E_{\rm {K \mbox{ }at \mbox{ } Ni(111)}}-{E_{\rm Ni(111)}-E_{\rm K}}$.}

\label{KonNiEnergie}
\begin{tabular}{cccccccc}
Site &  $d_{K-Ni1a}$  & $d_{Ni1b-Ni2}$ & $\delta$ & $d_{K-Ni \ nn}$ &
$E_{adsorption}$  \\
& [\AA] &  [\AA] &  [\AA] & [\AA] & $\left[\frac{E_h}{K \ atom}\right]$ \\
\\
\multicolumn{6}{c} {without rumpling}\\
fcc & 2.71 & 2.01  &   0    &  3.07   & -0.0532 \\
hcp & 2.70 & 2.01  &   0    &  3.06   & -0.0531 \\
bridge & 2.70 & 2.01 & 0    &  2.98   & -0.0531 \\
top & 2.76 & 2.01  &   0    &  2.76   & -0.0528\\
\\
\multicolumn{6}{c} {with rumpling}\\
fcc &       2.63 & 1.99  &  +0.07 &  3.06   & -0.0536\\
hcp &       2.63 & 1.99  &  +0.07 &  3.06   & -0.0535 \\
bridge &           2.65 & 1.97  &  +0.07 &  2.99   & -0.0535 \\
top &              2.68 & 1.93  &  +0.11 &  2.79   & -0.0544 \\
\\
\multicolumn{6}{c} {experiment}\\
exp. (LEED) \cite{Fisheretal,Kaukasoinaetal1993} & 2.70$\pm$0.04 & 
1.90$\pm$0.03 & 0.12$\pm$0.02  & 2.82$\pm$0.04 &  \\
exp. (SEXAFS) \cite{Adler1993} & & & & 2.92$\pm$0.02 \\
exp. (ARPEFS) \cite{Huangetal1993} & & 1.90$\pm$0.04 & 0.00$\pm$0.03 
& 3.02 $\pm$ 0.01 \\
exp. (PhD) \cite{Davis1994} & & $1.86 \pm 0.06$
& 0.01$\pm$0.09 &  2.87$\pm$0.06 \\
\end{tabular}
\end{center}
\end{table}

\begin{table}
\begin{center}
\caption{Mulliken spin and charge population for K/Ni(111), 
on different adsorption sites,
extracted from calculations with 3 nickel
layers. The population for the nickel atoms in
the individual layers is given. The left column corresponds to that atom
in the layer which is closest to the potassium adsorbate. For example,
in the top case, 
one atom in the first layer is closer than the other three atoms
resulting in two different values for the population. In the case
of the clean slab, all atoms within one layer are degenerate and thus
only one number is displayed. Also, when the populations were found to be
identical, only one number was displayed (usually from the second Ni layer
on, marked with 'all').}
\label{Kspintable}
\begin{tabular}{ccccccc}
site & 1$^{st}$ layer  & 2$^{nd}$ layer  & 3$^{rd}$ layer  &  K \\ \\
\multicolumn{5}{c}{spin}\\
fcc    & 0.58 (3 atoms) 0.62 (1 atom) & 0.68 (all) & 0.66 (all) & 0.001 \\
hcp    & 0.59 (3 atoms) 0.61 (1 atom) & 0.68 (all) & 0.66 (all) & 0.001 \\
bridge & 0.58 (2 atoms) 0.61 (2 atoms) & 0.68 (all)  & 0.66 (all) & 0.000 \\
top   & 0.55 (1 atom)0.61 (3 atoms)& 0.68 (3 atoms) 0.69 (1 atom) & 0.66 (all) & -0.003 \\
clean slab &   0.66  & 0.69 & 0.66 \\ \\
\multicolumn{5}{c}{charge}\\
fcc & 28.04 (3 atoms)28.00 (1 atom)& 28.10 (all) & 27.95 (all) & 18.71 \\
hcp & 28.04 (3 atoms)28.00 (1 atom)& 28.10 (all) & 27.95 (all) & 18.71 \\
bridge & 28.05 (2 atoms) 28.00 (2 atoms) & 28.10 (all) & 27.95(all) & 18.71 \\
top & 28.08 (1 atom) 28.00 (3 atoms) & 28.10 (all) & 27.95 (all) & 18.73 \\
clean slab & 27.95 & 28.09 & 27.95 & \\
\end{tabular}
\end{center}
\end{table}

\begin{table}
\begin{center}
\caption{Binding energies for Cl/Ni(111), 
extracted from calculations with 3 nickel
layers, as a function of the computational parameters. The geometry was
held fixed at the one optimized with the default parameters (table
\ref{ClonNitable}).}
\label{ClNiparametertest}
\begin{tabular}{ccccc}
site & default grid,  & 
better grid,  &
default grid,  & 
default grid,  \\
& default ITOL ($10^{-6}$,  $10^{-6}$) & default ITOL ($10^{-6}$,  $10^{-6}$)
& higher ITOL ($10^{-7}$,  $10^{-7}$) & 
even higher ITOL ($10^{-8}$,  $10^{-8}$)\\
fcc & -0.1358 & -0.1357 & -0.1359 & -0.1357 \\
hcp & -0.1351 & -0.1349 & -0.1338 & -0.1339 \\
bridge & -0.1325 & -0.1325 & -0.1317 & -0.1317 \\
top & -0.1177  & -0.1186 & -0.1180 & -0.1179 \\
\end{tabular}
\end{center}
\end{table}

\begin{table}
\begin{center}
\caption{Binding energies for K/Ni(111), 
extracted from calculations with 3 nickel
layers, as a function of the computational parameters. The geometry was
held fixed at the one optimized with the default parameters (table
\ref{KonNiEnergie}).}
\label{KNiparametertest}
\begin{tabular}{ccccc}
site & default grid,  & 
better grid,  &
default grid,  & 
default grid,  \\
& default ITOL ($10^{-6}$,  $10^{-6}$) & default ITOL ($10^{-6}$,  $10^{-6}$)
& higher ITOL ($10^{-7}$,  $10^{-7}$) & 
even higher ITOL ($10^{-8}$,  $10^{-8}$)\\
fcc &    -0.0536 & -0.0529 & -0.0534 & -0.0533 \\
hcp &    -0.0535 & -0.0528 & -0.0532 & -0.0531 \\
bridge & -0.0535 & -0.0528 & -0.0532 & -0.0532 \\
top &    -0.0544 & -0.0537 & -0.0539 & -0.0541 \\
\end{tabular}
\end{center}
\end{table}

\newpage

\newpage

\newpage
\begin{figure}
\caption{Band structure of ferromagnetic nickel, with the PBE functional.
The majority bands are drawn with full lines, the minority bands with dashed
lines.}
\label{nibulkbandstructure} 
\end{figure}
\centerline{\psfig{figure=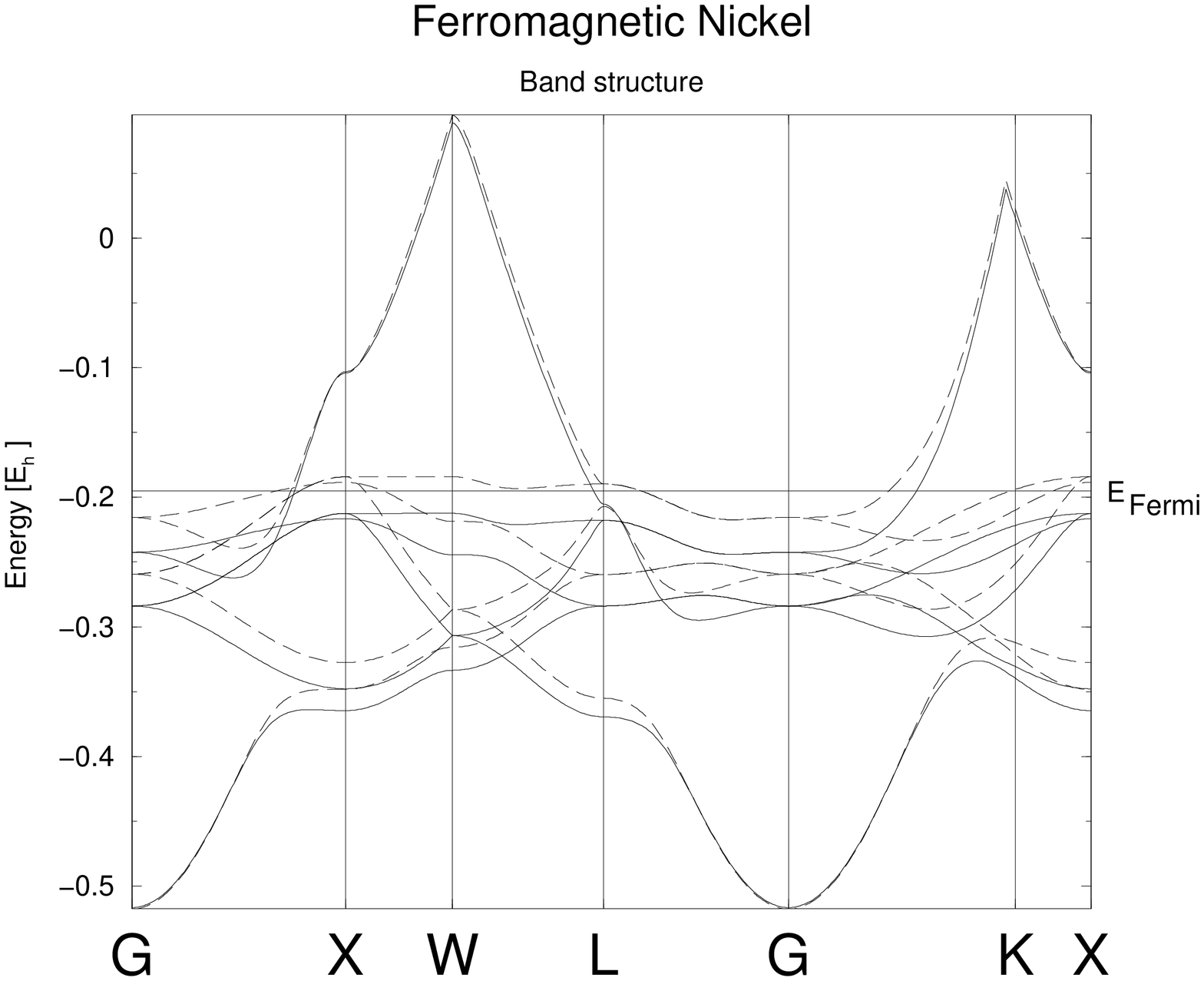,width=15cm,angle=270}}
\vfill

\newpage
\centerline{\psfig{figure=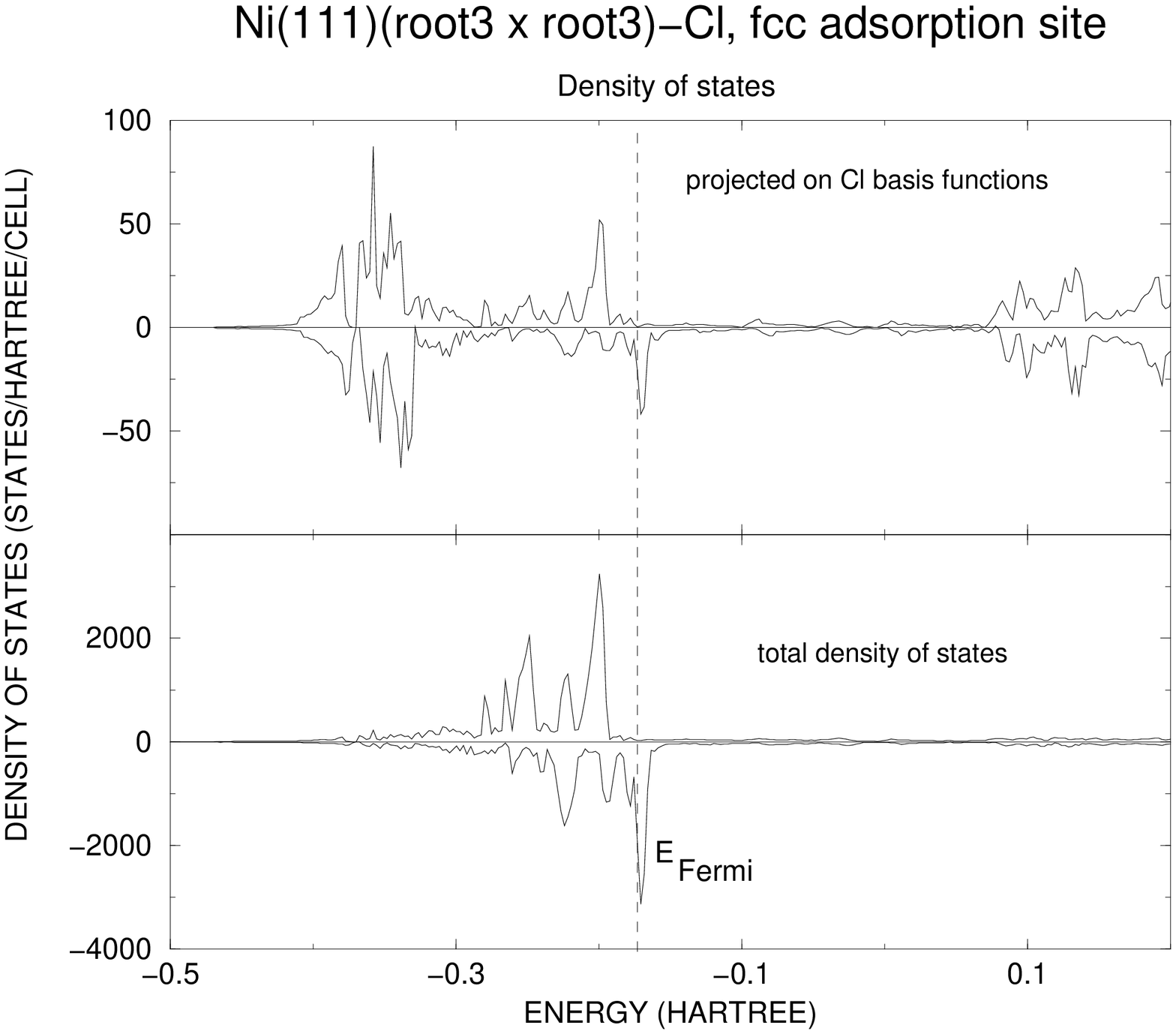,width=18cm,angle=270}}
\begin{figure}
\caption{The density of states, projected on the chlorine basis functions,
(upper two graphs); and projected
on all basis functions (lower two graphs), for majority (first and
third graph) and minority band (second and fourth graph).}
\label{ClNiDOS} 
\end{figure}
\vfill

\newpage
\centerline{\psfig{figure=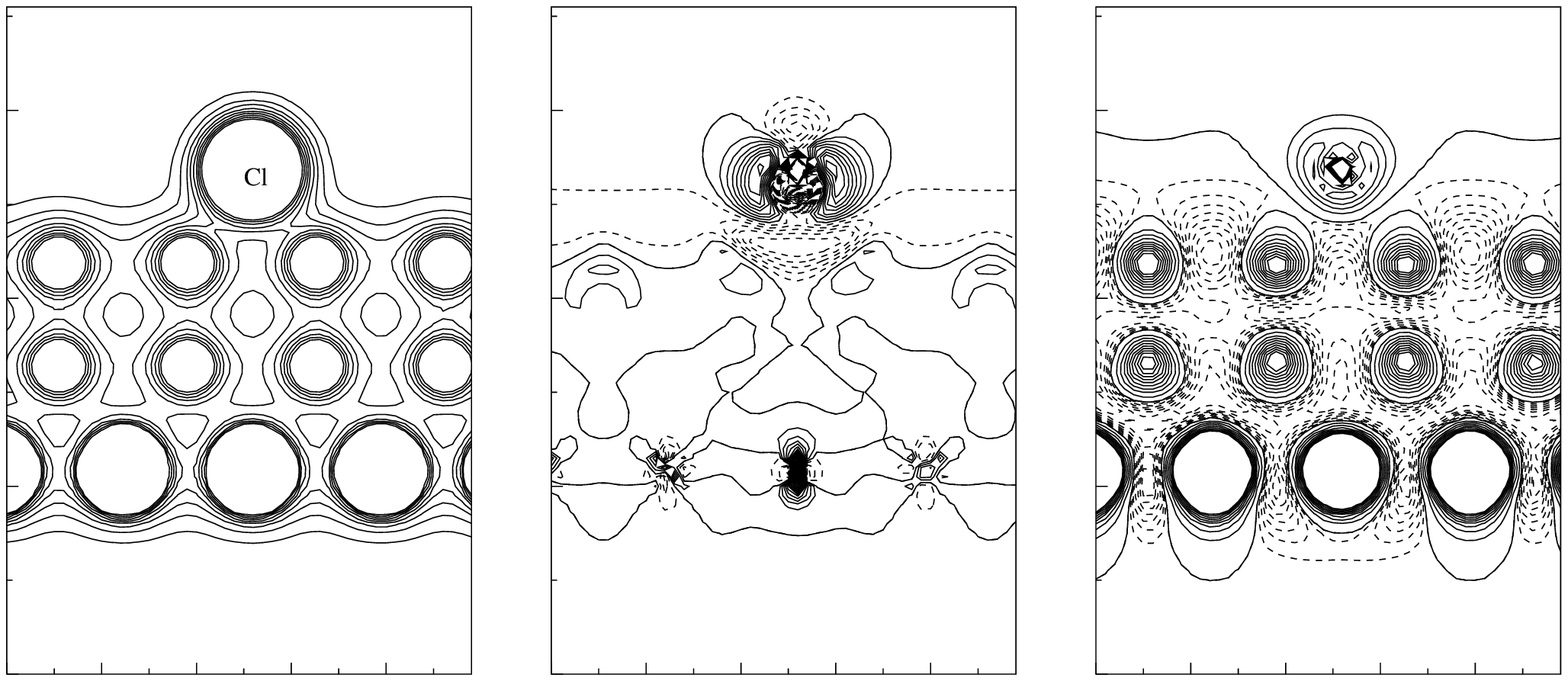,width=20cm,angle=90}}

\vspace{-6cm}
\begin{figure}
\caption{Charge density (left), charge density difference (middle)
and spin density (right) for Cl on Ni(111), 
from a three-layer slab, fcc site. 
The charge density is displayed with full lines, increasing in steps 
of 0.01, from 0 onwards. The charge density difference 
(charge of adsorbate system minus charge of a layer of chlorine atoms minus
charge of a clean nickel surface)
is in steps of 0.001, excess negative
charge is displayed with full lines (from 0 onwards), 
electron depletion with dashed lines (from -0.001 onwards).
Positive spin density is displayed
with full lines, increasing from 0 in steps of 0.001. Negative spin density
is displayed with dashed lines, decreasing from -0.0002
in steps of -0.0002. The
plane goes through the center of the Cl adsorbate, and through the 
nickel atom in the third layer vertically under the chlorine, and through
the nearest neighbors of this nickel atom in the third layer.}
\label{ClonNichargespindensity} 
\end{figure}
\vfill

\newpage
\centerline{\psfig{figure=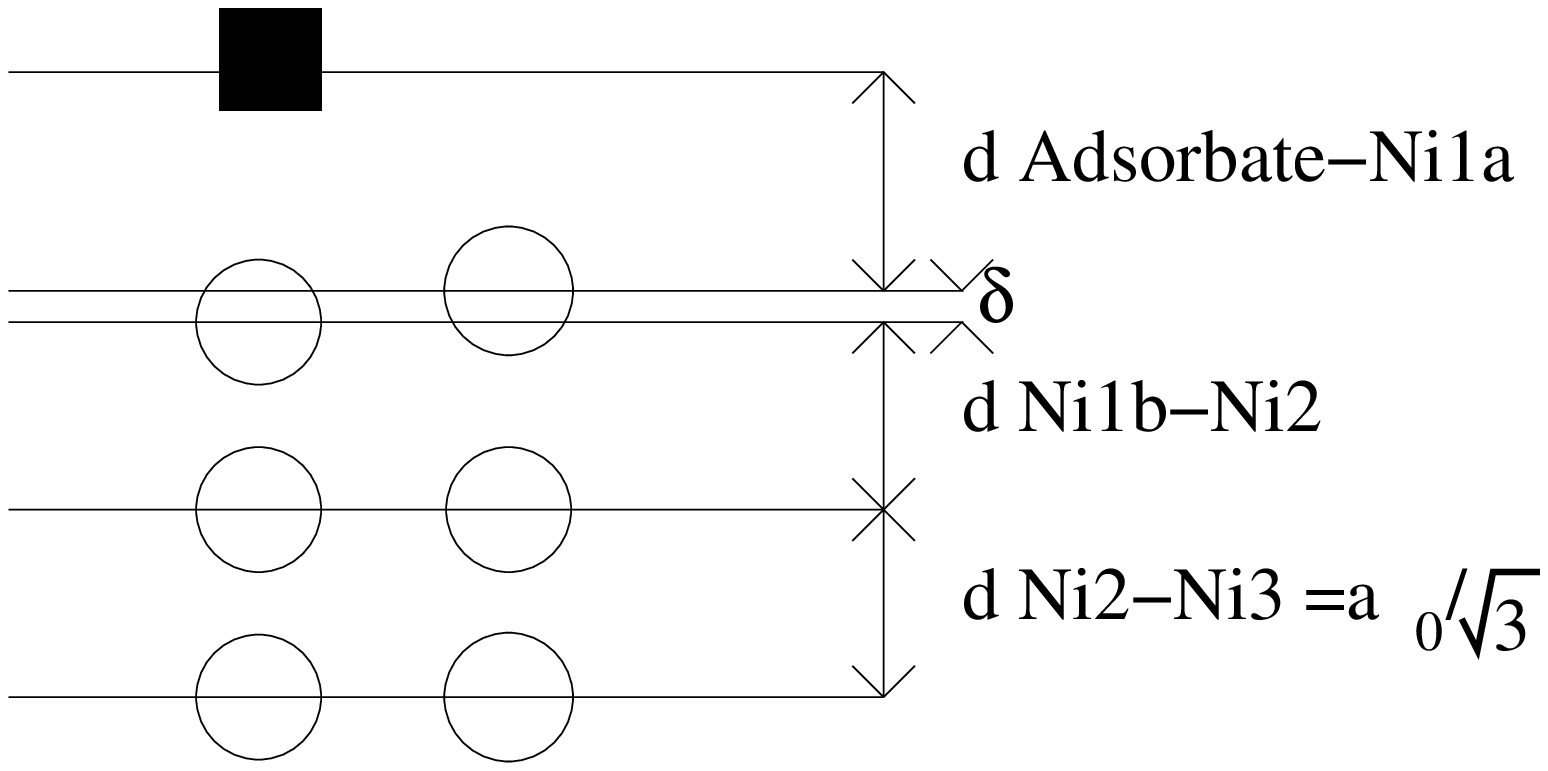,width=15cm,angle=270}}
\begin{figure}
\caption{Geometrical parameters for the adsorption studies. }
\label{geometricalparameters} 
\end{figure}
\vfill

\newpage
\centerline{\psfig{figure=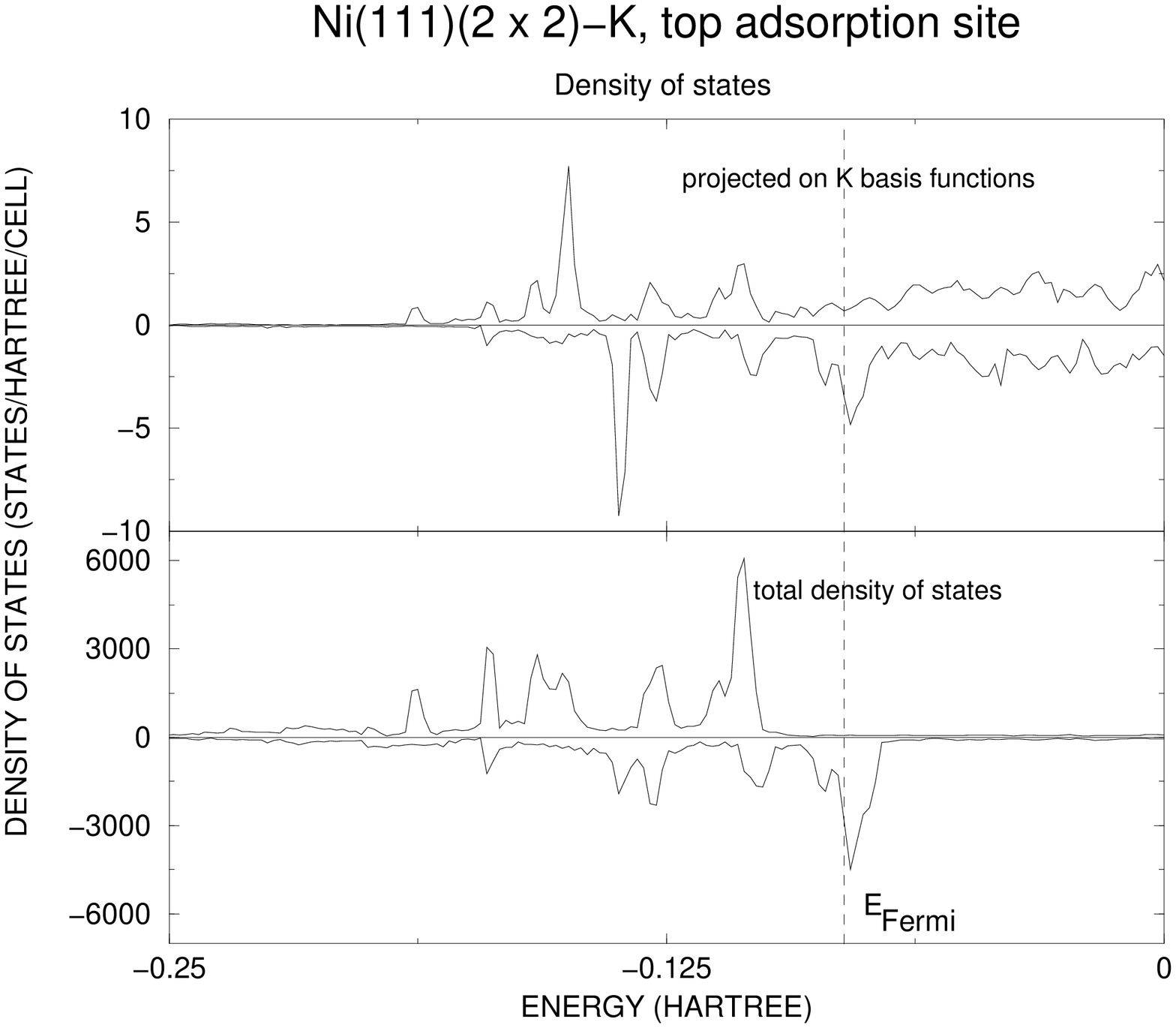,width=18cm,angle=270}}
\begin{figure}
\caption{The density of states, projected on the potassium basis functions,
(upper two graphs); and projected
on all basis functions (lower two graphs), for majority (first and
third graph) and minority band (second and fourth graph).}
\label{KNiDOS} 
\end{figure}
\vfill

\newpage
\centerline{\hspace{2cm}\psfig{figure=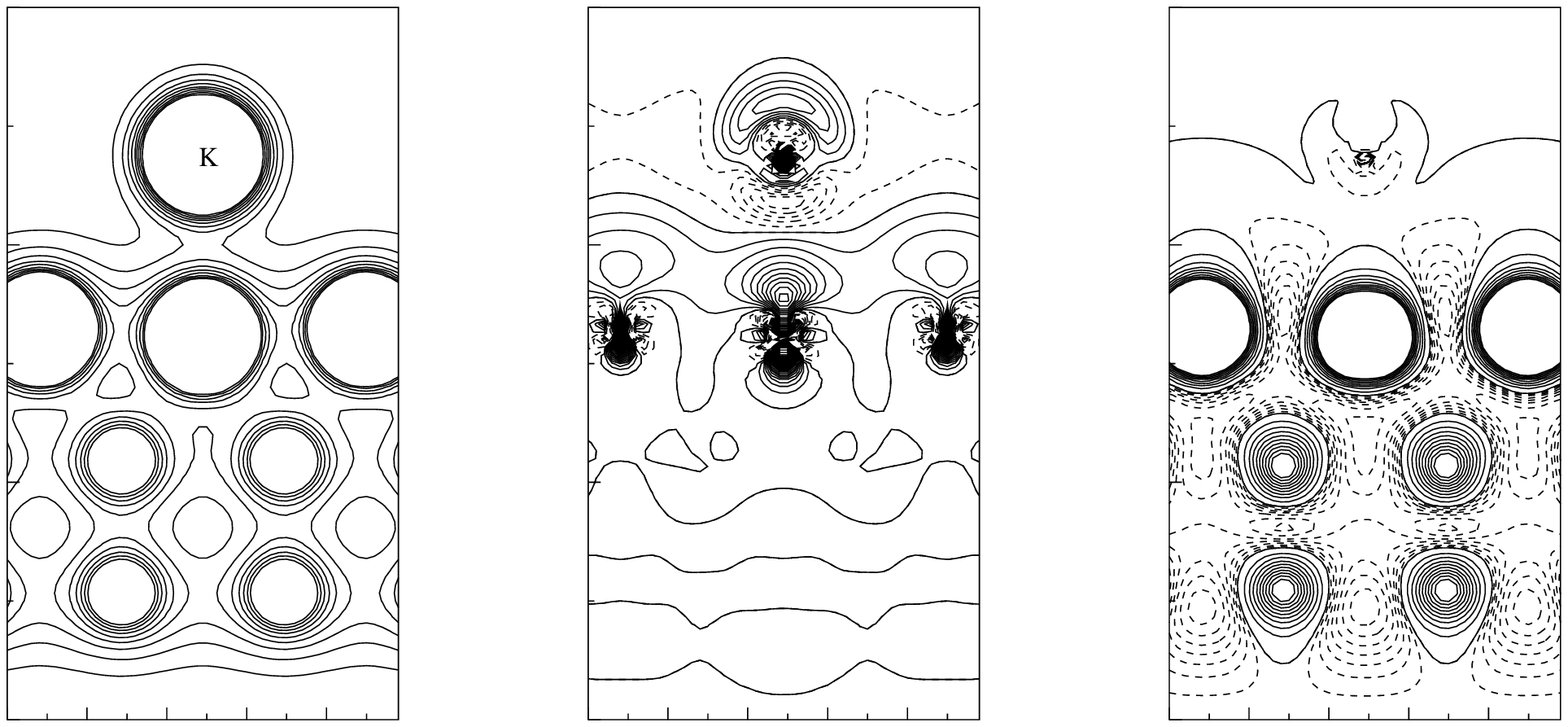,width=25cm,angle=90}}
\begin{figure}
\caption{
Charge density (left), charge density difference (middle)
and spin density (right) for K on Ni(111), 
from a three-layer slab, top site. 
The charge density is displayed with full lines, increasing in steps 
of 0.01, from 0 onwards. The charge density difference 
(charge of adsorbate system minus charge of a layer of potassium atoms minus
charge of a clean nickel surface)
is in steps of 0.001, excess negative
charge is displayed with full lines (from 0 onwards), 
electron depletion with dashed lines (from -0.001 onwards).
Positive spin density is displayed
with full lines, increasing from 0 in steps of 0.001. Negative spin density
is displayed with dashed lines, decreasing from -0.0002 
in steps of -0.0002. The
plane goes through the center of the K adsorbate, and through the 
nickel atom vertically below, and through
the nearest neighbors of this nickel atom in the first layer.
}
\label{KonNispin} 
\end{figure}
\vfill


\begin{references}
\bibitem{basisdetail} The energies of the neutral atoms were computed as 
follows: the outermost Cl $sp$-shells were 0.315 and 0.119
instead of 0.294 and 0.090, yielding an energy of -459.9385 $E_h$. For K,
three diffuse $sp$ shells (0.39, 0.21, 0.03) replaced the two used
for the adsorption studies (0.29, 0.08), yielding an energy of -599.6672
$E_h$. For the nickel atom, 7 additional even tempered diffuse functions
in the range from 0.01 to 0.25 ($sp$), and 5 from 0.016 to 0.6 ($d$),
were employed,
the $[5s4p1d]$ tight inner exponents were kept as in the bulk.
Thus, as a whole a $[12s11p6d]$ basis set was used. The PBE-energies
were -1507.933 $E_h$ $(d^8s^2)$; (PW:
-1508.311 $E_h$), and -1507.967 $E_h$ ($d^9s^1$); (PW: -1508.346 $E_h$).
The energies of the nickel atom were calculated with integer occupation
numbers, the $d_{z^2}$ and $d_{x^2-y^2}$ orbital were singly occupied
in the case of the $d^8s^2$ state, and the $d_{x^2-y^2}$ orbital in
the case of the $d^9s^1$ state.
\end{references}
\end{document}